\documentclass{ptapap}
\usepackage{subcaption}

\title{Study of slowly rotating CP stars observed with TESS}

\author{O. Kobzar}[1]
\author{V. Khalack}[1]
\author{D.Bohlender}[10]
\author{A. David-Uraz}[6]
\author{P. Kashko}[1,4]
\author{D.M. Bowman}[2]
\author{C. Lovekin}[3]
\author{D. Tvardovskyi}[1,5]
\author{M.Perron-Cormier}[1]
\author{E. Paunzen}[7]
\author{J. Sikora}[8,9]
\author{P. Lampens}[11]
\author{O. Richard}[12]

\affil[1]{D\'epartement de Physique et d'Astronomie, Universit\'e de Moncton, Moncton, N.B., Canada E1A 3E9}
\affil[2]{Instituut voor Sterrenkunde, KU Leuven, Celestijnenlaan 200D, B-3001 Leuven, Belgium}
\affil[3]{Department of Physics, Mount Allison University, Sackville, N.B., Canada E4L 1E6}
\affil[4]{Faculty of Physics, Taras Shevchenko National University of Kyiv, Kyiv, Ukraine 01033}
\affil[5]{Faculty of Mathematics, Physics and Information Technology, Odessa National I.I. Mechnikov University, Odessa, Ukraine 65000}
\affil[6]{Department of Physics \& Astronomy, University of Delaware, 217 Sharp Lab, Newark, DE 19716, USA}
\affil[7]{Department of Theoretical Physics and Astrophysics, Masaryk University, Kotl\'a\v{r}sk\'a 2, 611\,37 Brno, Czech Republic}
\affil[8]{Department of Physics, Engineering Physics \& Astronomy, Queen's University, Kingston, ON, Canada K7L 3N6}
\affil[9]{Department of Physics and Space Physics, Royal Military College of Canada, PO Box 17000 Kingston, ON, Canada K7K 7B4}
\affil[10]{National Research Council of Canada, Herzberg Astronomy and Astrophysics Research Centre, 5071 West Saanich Road, Victoria, BC, Canada V9E 2E7}
\affil[11]{Royal Observatory of Belgium, Ringlaan 3, B-1180 Brussels, Belgium}
\affil[12]{Laboratoire Universitaire et Particules de Montpellier, Université de Montpellier,  CNRS, Place E. Bataillon, 34095 Montpellier, France}

\begin{document}

\maketitle

\begin{abstract}
Since the end of 2018, the Transiting Exoplanet Survey Satellite (\textit{TESS})
provides high-quality space data on stellar photometry to the astronomical community. We present the results of an analysis of \textit{TESS} photometric data for known slowly rotating magnetic chemically peculiar (mCP) stars.
In general, mCP stars show an inhomogeneous distribution of elements in their
stellar atmospheres that leads to spectroscopic (line profile) and photometric
(light curve) variations over the rotation period.
In the frame of the oblique magnetic rotator (OMR) model, patches of enhanced chemical abundance on the stellar surface reveal the frequency of stellar rotation. Using this approach, we have compiled a list of slowly rotating mCP stars with rotation periods longer than two days from the analysis of the photometric data provided by \textit{TESS} for the first eight sectors of observations. Slowly rotating mCP stars usually possess a hydrodynamically stable stellar atmosphere where a magnetic field can amplify the process of atomic diffusion and leads to the horizontal and vertical stratification of chemical abundances.

\end{abstract}

\section{Introduction}

Chemically peculiar (CP) stars on the upper main sequence are identified by the presence of enormously strong or weak absorption lines in their spectra for certain chemical elements. \citet{Preston1974} divided the known CP stars into four classes according to their peculiarity type and the presence of a magnetic field. Among them the most interesting class is the class of ApBp stars which show horizontal (presence of overabundance patches) and
vertical abundance stratification \citep{Khalack2017}, and possess a significant global magnetic field \citep{Bychkov2003,Buysschaert2018b}.
They usually are known as the magnetic CP (mCP) stars.

To explain the chemical peculiarities observed in CP stars, \citet{Michaud1970} proposed the mechanism of atomic diffusion which also takes into account the competition between radiative and gravitational forces.
In ApBp stars, the presence of a magnetic
field can stabilize the turbulent motions in the stellar atmosphere and amplify the
atomic diffusion of specific chemical elements \citep{Alecian2010}. Thus, the stratification of chemical species in different parts of the stellar atmosphere of ApBp stars depends on the intensity and structure of the local magnetic field.

Based on statistical analysis, \citet{Gomez1998} showed that only 10-15 \% of stars on upper main sequence are CP stars, with a significant fraction of them possessing a magnetic field of predominantly dipolar structure. Taking into account that the rotation axis usually does not coincide with the axis of the magnetic dipole,  \citet{Stibbs50} proposed the oblique magnetic rotator (OMR) model to describe the variability of the mean longitudinal magnetic field observed in ApBp stars. 
One can use photometric observations to detect LC variability caused by the presence of co-rotating overabundance patches in the stellar atmosphere of mCP stars \citep{Bowman2018b,David-Uraz2019,Sikora2019}. According to the OMR model \citep{Stibbs50}, such photometric variations may reveal the frequency of stellar rotation and its first harmonic.

\section{\textit{TESS} data collection}

The photometric observations recently obtained by the Transiting Exoplanet Survey Satellite (\textit{TESS}) were used to identify, from a sample of relatively bright ApBp stars for which spectral observations and magnetic measurements are available, those that have a long (> 2 days) rotational period (see Table~\ref{tab1}). A majority of stellar parameters for these stars were collected from the \textit{TESS} Input Catalogue (TIC) \citep{Stassun2019a}. The \textit{TESS} data were downloaded via the Mikulski Archive for Space 
Telescopes\footnote{https://archive.stsci.edu/tess/bulk\_downloads} 
and are publicly available. The extracted flux measurements were transformed into time series of stellar magnitudes with timestamps in units of Barycentric Julian Date (BJD).

From the analysis of the \textit{TESS} light curves for the first eight sectors, we compiled a sample of approximately 550 candidates that show a frequency peak which could be the stellar rotation frequency and its first harmonic. To carry out such analysis, we used the \textit{TESS}-AP automatic procedure \citep{Khalack2019} that consists of several codes including Period04 \citep{Period04} designed for automatic data analysis. From this sample we selected eight slowly rotating (P > 2 days) and relatively bright (V < 8.0~mag) mCP stars that have small v$\sin{i}$ (see Table~\ref{tab1}) to compare their phased light curve (LC) with the phase curve of the mean longitudinal magnetic field measurements $<B_{\rm z}>$. Five of the selected stars are discussed here.

\begin{table*}
\begin{center}
\caption{List of properties of studied CP stars }
\label{tab1}
\def\arraystretch{0.98}
\setlength{\tabcolsep}{3pt}
\begin{tabular}{lrcccc}
\hline
Name&{$T_{\rm eff}$ (K)} &{$\log{g}$}&Period (d)&{v$\sin{i}$ (km~s$^{-1}$)}&{v$_{\rm r}$  (km~s$^{-1}$)}\\

HD &TIC &TIC & This study & SIMBAD & SIMBAD\\
\hline
10840 & $11600 \pm 500$& $3.60 \pm 0.20$& $2.09765 \pm 0.00003$& $35.0 \pm 5.0$& $19.4 \pm 2.1$\\
22920 & $13640 \pm 200$& $3.65 \pm 0.10$& $3.946   \pm 0.001$  & $37.0 \pm 5.0$& $18.0 \pm 4.0$\\
24712 & $7280  \pm 210$& $4.11 \pm 0.34$& $12.44   \pm 0.02$   & $18.0 \pm 0.0$& $23.2 \pm 0.4$\\
38170 & $10000 \pm 260$& $           - $& $2.7664 \pm 0.0005$  & $65.0 \pm 9.0$& $36.3 \pm 0.6$\\
63401 & $13500 \pm 500$& $4.20 \pm 0.20$& $2.4149 \pm 0.0004$  & $52.0 \pm 4.0$& $22.0 \pm 1.4$\\
74521 & $10790 \pm 500$& $3.47 \pm 0.30$& $7.037   \pm 0.003$  & $18.0 \pm 2.0$& $27.5 \pm 1.4$\\
77314 & $9670  \pm 250$& $3.83 \pm 0.48$& $2.864  \pm 0.001$   & $   -        $& $  -         $\\
86592 & $9100  \pm 240$& $4.35 \pm 0.31$& $2.8886  \pm 0.0003$ & $16.0 \pm 2.0$& $12.7 \pm 0.3$\\
\hline
\end{tabular}
\end{center}
\end{table*}

\section{Results for individual stars}

\subsection{HD~22920 (TIC~301621458 = FY~Eri)}

 We determined the rotation period $P = 3.946 \pm 0.001$~d for HD~22920, which is in accordance with the results published by \citet{Bartholdy88}. Using the OMR model, we found a weak correlation of the phased $<B_{\rm z}>$ measurements with the LC phase diagram derived for this star. We assume that the magnetic field amplifies atomic diffusion in stellar atmosphere of HD~22920 and therefore influences the distribution of the chemical elements \citep{Alecian2010}. The values of $T_{\rm eff} = 13640 \pm 200$~K, $\log{g} = 3.72 \pm 0.20$, radial velocity v$_{\rm r} = 18.6 \pm 4.5$~km~s$^{-1}$, and v$\sin{i} = 37.8 \pm 5.4$~km~s$^{-1}$ were estimated by  \citet{Khalack+Poitras15}.

\subsection{HD~24712 (TIC~279485093 = DO~Eri)}

From the Fourier analysis of the \textit{TESS} data obtained for HD~24712, we detected a 
signal at the frequency and its first harmonic that may correspond to a rotation  period of $P = 12.44 \pm 0.02$~d. We can also confirm that HD~24712 shows roAp type pulsations with periods around 6~min. 

\subsection{HD~63401 (TIC~175604551 = OX~Pup)}

The atmospheric stellar parameters, $T_{\rm eff} = 13360 \pm 200$~K and $\log{g} = 4.1 \pm 0.2$, were obtained by fitting the Balmer line profiles in the available ESPaDOnS spectra by \citet{Kashko2019}. The frequency of the identified signal may be interpreted as the frequency corresponding to the rotation period $P = 2.4149 \pm 0.0004$~d. HD~63401 was observed in two sectors (approximately 55 days) as well as HD~10840 and HD~38170. These additional data empower to improve the value of period and its estimation error. It seems that the LC phase diagram is weakly correlated with the phased $<B_{\rm z}>$ measurements. Considering the OMR model this would imply that HD~63401 has overabundance patches close to the magnetic poles.

%

\begin{figure}
\centering
\includegraphics[width=3.0in, angle=-90]{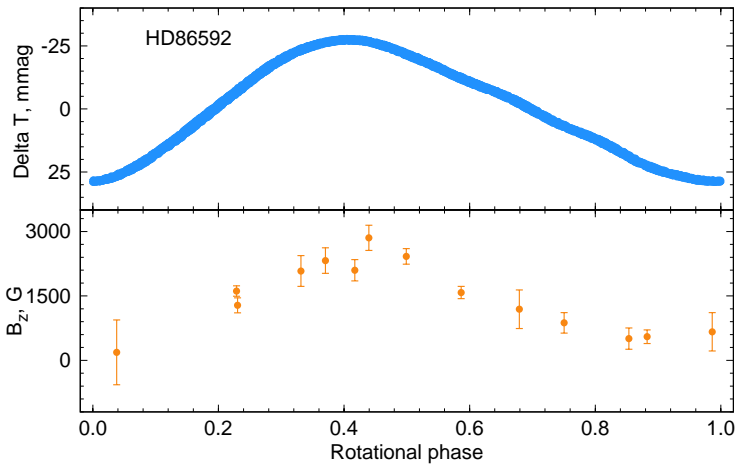}
\caption{Phased light curve (LC) (top) and phased curve of the longitudinal magnetic field measurements (bottom) for HD~86592}
\label{fig:1}
\end{figure}

\subsection{HD~77314 (TIC~270487298 = NP~Hya)}
 
We determined the period $P = 2.864\,\pm\,0.001$~d, which may coincide with the rotation period. The measurements of the mean longitudinal magnetic field $<B_{\rm z}>$ were carried out by Bohlender with dimaPol on the DAO 1.8m telescope. We assumed that the magnetic field has an important influence on the stratification of elements in the stellar atmosphere of HD~77314. In this case, the OMR model can explain the LC variability due to presence of horizontal abundance inhomogeneities.

\subsection{HD~86592 (TIC~332654682 = V359~Hya)} 
\label{HD86592}

A stellar rotation period of $P = 2.8886\,\pm\,0.0003$~d was derived for HD~86592 from the analysis of its light curve and used to build the phased light curve (see top panel at Fig.~\ref{fig:1}). This value coincides with the period reported by \citet{Babel1997}.  We used the dimaPol mean longitudinal magnetic field measurements obtained by Bohlender to compare the variability of $<B_{\rm z}>$ with the LC phase diagram. We found that the data are correlated (Pearson's correlation coefficient: $R = -0.95$). Considering the OMR model, one can see that the maximum visibility of the positive magnetic pole corresponds to the maximum on the LC phase diagram (see Fig.~\ref{fig:1}).

\section{Discussion}

This work was carried out within the framework of the MOBSTER Collaboration \citep{David-Uraz2019,Sikora2019} and the VeSElkA project \citep{Khalack2015, Khalack_new2019}. 
In this study, we identified the targets that show photometric variability which may be attributed to stellar rotation and for which high-resolution spectra and measurements of magnetic field are available. 
The light curves of the analysed mCP stars are used to reveal the frequency of stellar rotation and its first harmonic.
This variable behaviour can be explained in terms of co-rotating overabundance spots present in the stellar atmosphere of these stars. Taking into account that, in some cases, the variability of LC and of mean longitudinal  magnetic field have the same period, and their phase diagrams are correlated (see subsection \ref{HD86592}); we can apply the OMR model \citep{Stibbs50} to describe them.
Consequently, we assume that the detected frequency peak in the LC and its first harmonic corresponds to the stellar rotational frequency.

\section{Acknowledgments}

V.K., C.L. and A.D.U. acknowledge the support from the Natural Sciences and Engineering Research Council of Canada (NSERC). O.K. and V.K. are thankful to the Facult\'{e} des \'{E}tudes Sup\'{e}rieures et de la Recherche and to the Facult\'{e} des Sciences de l'Universit\'{e} de
Moncton for the financial support of this research. P.K. and D.T. acknowledge the support from the Globalink Research Internships program. This paper includes data collected by the \textit{\textit{TESS}} mission. The research leading to these results has received funding from the European Research Council (ERC) under the European Unions Horizon 2020 research and innovation programme (grant agreement No. 670519: MAMSIE). The study is based on observations obtained at the Dominion Astrophysical Observatory, Herzberg Astronomy and Astrophysics Research Centre, National Research Council of Canada.
This research has made use of the SIMBAD database, operated at CDS, Strasbourg,
France. Some of the data presented in this paper were obtained from the Mikulski
Archive for Space Telescopes (MAST).
STScI is operated by the Association of Universities for Research in Astronomy,
Inc., under NASA contract NAS5-2655.

\bibliographystyle{ptapap}
\bibliography{kobzar}
\end{document}